\documentclass{ws-ijmpd}

\usepackage{graphicx}

\newcommand{\be}{\begin{equation}}
\newcommand{\eea}{\end{eqnarray}}

\begin{document}

\markboth{Ray, Das, Rahaman \& Ray} 
{Physical properties of Tolman-Bayin solutions}

\catchline{}{}{}{}{}
%
%%%%%%%%%%%%%%%%%%%%%%%%%%%%%%%%%%%%%%%%%%%%%%%%%%%%%%%%%%%%%%%%%%%%
\title{Physical properties of Tolman-Bayin solutions: \\
some cases of static charged fluid spheres in general relativity}

\author{Saibal Ray\footnote{Associate, IUCAA, Pune 411 007, India}}
\address{Department of Physics, Barasat Government College, \\
North 24 Parganas, Kolkata 700 124, India\\
saibal@iucaa.ernet.in}

\author{Basanti Das}
\address{Belda Prabhati Balika Vidyapith, Belda, \\
West Midnapur 721 424, W.B., India}

\author{Farook Rahaman}
\address{Department of Mathematics, Jadavpur University, \\
Kolkata 700 032, W.B., India}

\author{Subharthi Ray}
\address{Inter-University Centre for Astronomy and Astrophysics, \\
Post Bag 4, Ganeshkhind, Pune 411 007, India\\
sray@iucaa.ernet.in}

\maketitle

%\begin{history}
%\received{Day Month Year}
%\revised{Day Month Year}
%\comby{Managing Editor}
%\end{history}

%%%%%%%%%%%%%%%%%%%%%%%%%%%%%%%%%%%%%%%%%%%%%%%%%

\begin{abstract}
In this article, Einstein-Maxwell space-time has been considered
in connection to some of the astrophysical solutions as previously
obtained by Tolman (1939) and Bayin (1978). The effect of
inclusion of charge into these solutions has been investigated
thoroughly and also the nature of fluid pressure and mass density
throughout the sphere have been discussed. Mass-radius and
mass-charge relations have been derived for various cases of the
charged matter distribution. Two cases are obtained where perfect
fluid with positive pressures give rise to {\em electromagnetic
mass} models such that gravitational mass is of purely
electromagnetic origin.
\end{abstract}

\keywords{general relativity; charged dusts; charged stars.}

\section{Introduction}
The so-called Schwarzschild interior solutions obtained by using
Einstein's field equations corresponding to various spherically
symmetric static perfect fluid distributions usually suffer from
the well-known problem of singularity. An uncharged
incompressible fluid sphere of mass $m$ cannot be held in
equilibrium below certain radius $a = 9m/4$ and even demands a
larger value for $a$ related to physically reasonable equation of
state~\cite{buc59}. One way to overcome this singularity due to
gravitational collapsing of a spherically symmetric material
distribution is to include charge to the neutral case. It is
observed that gravitational collapse can be averted in the
presence of charge where gravitational attraction is counter
balanced by the electrical repulsion in addition to the pressure
gradient~\cite{fel95yf,sha01mm,iva02} 

But even then many questions came up regarding the stability of
the charged sphere and also about the amount of charge.
Bonnor~\cite{bon65} worked on this type of charge inclusion model
and showed that a dust cloud of arbitrarily large mass and small
radius can remain in equilibrium if it has an electric charge
density related to the mass density by $\sigma = \pm \rho$.
According to Stettner~\cite{ste73}, a fluid sphere of uniform
density with a net surface charge is more stable than without
charge. However, Glazer~\cite{gla76} by considering radial
pulsations showed that Bonnor~\cite{bon65} model is electrically
unstable. He also explicitly established the effects of electric
charge upon dynamical stability~\cite{gla79}. The work of Whitman
and Burch~\cite{whi81b}, for arbitrary charge and mass distribution
showed how charged analogue cases give more stability. They also
applied the pulsation equations to the charged solution of Pant
and Sah~\cite{pan79s}, which represents the charged analogue of
Tolman~\cite{tol39} solution of type VI, and got unsatisfactory
results in connection to the boundary condition which is
incompatible with densities and pressures. de Felice et
al.~\cite{fel95yf} proposed a model for charged perfect fluid and
concluded that inclusion of charge inhibits the growth of
space-time curvature which has a great role to avoid
singularities. However, very recently Ray et al.~\cite{ray03x} and
Ghezzi~\cite{ghe05} have studied the effect of electric charge on
compact stars assuming the charge distribution is proportional to
the mass density. Both the group have argued that with the huge
amount of charge and strong electric field these type of stars
would be unstable and eventually would form charged black holes.

Therefore, starting from the very well-known
Weyl-Majumdar-Papapetrou~\cite{wey17,maj47,pap47}
type general relativistic charged dust solutions, via
Bonnor~\cite{bon65}  and several others (some of which have been
mentioned in the above introductory discussions), one can arrive
at the realm of charged compact stellar objects. But a genuine
question can be raised in this context that: what about the
normal stars, do they carry electric charge and if at all, what
will be the stability of their configurations? As such regarding
the effect of charge and possibility of holding charge by the
stars are not unavailable in the past literature~\cite{pan22,ros24,edd26,cow29}
. As a
continuation of such study Shvartsman~\cite{shv71} argued that while
astrophysical systems are usually thought to be electrically
neutral, this may not be always true in the real situation. His
analysis is based on the exchange processes between stars and the
surrounding medium. Very recently, Neslu{\u{s}}an~\cite{nes01} has
reminded about the existence of global electrostatic field of the
sun and other normal stars. He has given a general charge-mass
relation $q_r = [2\pi \epsilon _0 G(m_p - m_e)m_r]/q$ where
$\epsilon _0$ is the primitivity of vacuum, $G$ is the
gravitational constant, $m_r$ is the stellar mass in the sphere
of radius $r$, $m_p$ and $m_e$ are respectively the mass of
proton and electron having charge $q$, while $q_r$ is the global
electrostatic charge inside the star. He got it to be $q_r =
77.043 m_r$ when $q_r$ in Coulomb and $m_r$ in solar masses
corresponding to an ideally quiet, perfectly spherical,
non-rotating star.

So, motivated by all the above facts regarding the charged sphere
in connection to normal stars, we have considered here charge
analogue of Tolman VI~\cite{tol39} and Bayin~\cite{bay78} type
solutions which represent very important class of astrophysical
solutions. However, in a series of work Ray and
Das~\cite{ray02d,ray04d,ray05d} have already been studied some
specific aspects of these type of solutions in different context.
In the present work we analyze different
physical properties of the static charged stellar model. We show
that some cases provide {\em electromagnetic mass model} (EMMM)
where all the physical parameters, including the gravitational
mass, vanish due to vanishing charge
density~\cite{lor04,fey64ls}. It has also shown that EMMM do
exists even with positive pressure which clearly contradicts the
observation done by Ivanov~\cite{iva02} that EMMM are allways
associated with repulsive pressure. The effect of charge
inclusion in these EMMM, along with other solutions, has been
investigated thoroughly in connection to mass-radius and
mass-charge relations. The nature of fluid pressure and mass
density throughout the fluid sphere have also been discussed.

\section{The Einstein-Maxwell field equations}
Let us consider a static spherically symmetric matter
distribution corresponding to the line element
\begin{eqnarray}
ds^{2} = A^{2} dt^{2} - B^{2} dr^{2} - r^{2} (d\theta^{2} + sin^{2}\theta
d\phi^{2})
\label{metric}
\end{eqnarray}
where $A$ and $B$ are function of the radial coordinate $r$ only.

Then the set of Einstein-Maxwell field equations, in the co-moving
coordinates, for the above line element may be explicitly written~\cite{ray02d} as
\begin{eqnarray}
\frac{1}{B^2}\left(\frac{2B'}{Br} - \frac{1}{r^2}\right) +
\frac{1}{r^2} = 8\pi\rho + \frac{q^2}{r^4},
\end{eqnarray}
\begin{eqnarray}
\frac{1}{B^2}\left(\frac{2A'}{Ar} + \frac{1}{r^2}\right) -
\frac{1}{r^2} = 8\pi p - \frac{q^2}{r^4},
\end{eqnarray}
\begin{eqnarray}
\frac{1}{B^2}\left[\frac{A''}{A} - \frac{A'B'}{AB} + \frac{1}{r}
\left(\frac{A'}{A} - \frac{B'}{B}\right)\right]
= 8\pi p + \frac{q^2}{r^4},
\end{eqnarray}
where the total charge within a sphere of radius $r$, in terms of
the 4-current $J^i$, can be given by
\begin{eqnarray}
q = 4\pi \int_0^r J^0 r^2 AB dr.
\end{eqnarray}

\section{The solutions to the field equations}
\subsection{Bayin's class of solution}
Now, assuming $A^{\prime}/Ar=C(r)$ and then equating~(3) and (4),
we get Bernoulli equation for B(r) and C(r) as follows
\begin{eqnarray}
\frac{dB}{dr} = \left[\frac{1 - \frac{2q^2}{r^2}}{\left(C +
\frac{1}{r^2}\right)r^3}\right]B^3 + \left[\frac{C^{2}r - \frac{1}{r^3} +
\frac{dC}{dr}}{C + \frac{1}{r^2}}\right]B.
\end{eqnarray}
Therefore, choosing $q(r) = Kr^n$ we get the solutions for $p(r)$
and $\rho(r)$ corresponding to different values of the parameter
$n$ (for detail results vide ref.~\cite{ray02d}).

{\bf Case I: For $n = 1$}
%\begin{widetext}
\begin{eqnarray}
\nonumber
8\pi p &=&\frac{K^2}{r^2} - \frac{1}{r^2} + \left\{\frac{a_0 + 3a_1
r}{(a_0 + a_1 r)r^2}\right\} \times \\
\nonumber
&& \left\{1 +B_0 r^2 - \frac{2r}{C_0}
 + \frac{4r^2}{C_0 ^2}\ln\frac{D}{r} - K^2 \right. \\
&& \left.  - 4K^2 r^2\left(\frac{2}{{C_0}^2}\ln\frac{D}{r}
 - \frac{1}{C_0 r} + G_1 \right)\right\}, \label{bay11p}
\end{eqnarray}
%\end{widetext}
%\begin{widetext}
\begin{eqnarray}
\nonumber
8\pi\rho &=&\frac{K^2}{r^2} + \frac{4}{C_0}\left(\frac{1}{r} +
\frac{1}{D}\right) - 3B_0 +\frac{4K^2}{rD}  - \frac{12}{C_0
^2}\ln\frac{D}{r} \\
&& -8K^{2}r\left(\frac{2}{C_0 ^2}\ln\frac{D}{r} -
\frac{1}{C_0 r} + G_1 \right). \label{bay11rho}
\end{eqnarray}
%\end{widetext}
where $C_0$ and $G_1$ are the constants of integration. Here and
in what follows $D$ has been substituted for $(C_0 + 2r)$.\\

 {\bf Case I: For $n = 3$}

%\begin{widetext}
\begin{eqnarray}
\nonumber
8\pi p &=& K^2 r^2 - \frac{1}{r^2} + \frac{a_0 + 3a_1 r}{(a_0 + a_1
r)r}\left[1 + B_0 r^2 - \frac{2r}{C_0} \right. \\
\nonumber
&& \left. +\frac{4r^2}{C_0^2}\ln\frac{D}{r} - 2Kr^4 - \frac{K^2
r^2}{2}\left(\frac{D^2}{2} - 2C_0 D \right. \right. \\
&& \left. \left. + C_0^2\ln D + G_2 \right)\right],
\label{bay13p}
\end{eqnarray}
%\end{widetext}
%\begin{widetext}
\begin{eqnarray}
\nonumber
8\pi\rho &=& - 11K^2 r^2 + \frac{4}{C_0}\left(\frac{1}{r} +
\frac{1}{D}\right) \\
\nonumber
&& - 3B_0 + \frac{4K^2 r^3}{D}  - \frac{12}{C_0
^2}\ln\frac{D}{r} \\
&& - K^2\left[\frac{D^2}{2} + 2C_0(C_0 + r) + C_0
^2\ln D + 8G_2\right]. \label{bay13rho}
\end{eqnarray}
%\end{widetext}
where $G_2$ is the constant of integration.\\

{\bf Case II: For $n = 1$}

%\begin{widetext}
\begin{eqnarray}
\nonumber
8\pi p &=& \frac{1}{W_0 ^2}\left[2(C_1 r^2 + W_0 ^2 - r^4 - 2W_0 ^2
K^2)^{1/2} \right. \\
&& \left. - r^2 + C_1\right] -\frac{K^2}{r^2},
\label{bay21p}
\end{eqnarray}
%\end{widetext}
\begin{eqnarray}
8\pi\rho = \frac{1}{W_0 ^2}(5r^2 - 3C_1) + \frac{K^2}{r^2}.
\label{bay21rho}
\end{eqnarray}
where $C_1$ is the constant of integration.\\

{\bf Case II: For $n = 3$}

%\begin{widetext}
\begin{eqnarray}
\nonumber
8\pi p &=& \frac{1}{W_0 ^2}\left[2\left\{C_1 r^2 + W_0 ^2 - (1 -
2W_0 ^2 K^2)r^4\right\}^{1/2} \right. \\
&& \left. -r^2 +C_1\right] + 3K^2 r^2,
\label{bay23p}
\end{eqnarray}
%\end{widetext}
\begin{eqnarray}
8\pi\rho = \frac{1}{W_0 ^2}(5r^2 - 3C_1) - 11K^2 r^2.
\label{bay23rho}
\end{eqnarray}

{\bf Case III: For $n = 1$}

\begin{eqnarray}
8\pi p = \frac{1}{r^2} \left[K^2 - 1 + (2K^2 - 1) (C_3 r - 2) C_3 r \right],
\end{eqnarray}
\begin{eqnarray}
8\pi\rho = \frac{1}{r^2} \left[\frac{1 - K^2}{r^2} - \frac{2C_3 r
+ (2K^2 - 1)(C_3 r - 2)^3}{(C_3 r - 2)r^2} \right].
\end{eqnarray}
where $C_3$ is the constant of integration.

\subsection{Tolman VI type class of solution}

In their work Ray \& Das~\cite{ray02d} considered the solution of
Pant \& Sah~\cite{pan79s} to discuss about the electromagnetic
mass model (EMMM) while in another paper \cite{ray04d} they have
discussed the role for equation of state regarding the EMMM.
However, in the present work we would like to find out some other
features of the solutions, specially the condition to have
positive pressure. The solutions given by Pant \& Sah~\cite{pan79s}
corresponding to the metric (Eq. \ref{metric}),
where $e^{\nu/2} = A$, $e^{\lambda/2} = B$ and $K=E^2/8\pi$ in the
present mathematical nomenclature, become
\begin{eqnarray}
\rho = \frac{1}{16\pi r^2}[1 - c(n - 1 )^2], \label{tol17rho}
\end{eqnarray}
\begin{eqnarray}
p = \frac{1}{16\pi r^2}[c(n+1)^2 -1] \label{tol18p}
\end{eqnarray}
where
\begin{eqnarray}
c = \left[1 - \frac{2m}{a} + \frac{{q}^2}{{a}^2}\right] = \left[1 -
\frac{{2q}^2}{{a}^2}\right]( 1 + 2n - n^2)^{-1}.
\end{eqnarray}
The above set of solutions, in view of $c$, with $\Lambda=0$ and
$B=0$ represents the charged analogue of Tolman's \cite{tol39}
solution VI and thus in the absence of the total charge $q$
reduces to the neutral one (the sub-case $C$ of uncharged fluid
sphere in Pant \& Sah~\cite{pan79s}). The corresponding total
gravitational mass $m(r=a)$, can be obtained as
\begin{eqnarray}
m = \frac{na^2(2 - n) + 2q^2}{2(1+2n-n^2)a}.
\end{eqnarray}

\section{Detailed study of some of the solutions}

\subsection{Bayin's solutions}

\noindent{\bf Case I: For $n = 1$}

From the expressions for pressure $p(r)$ and density $\rho(r)$, as
given in Eqs. (\ref{bay11p}) and (\ref{bay11rho}), for the central
values, viz., $r = 0$, we have $p_0/\rho_0 = -1 $ which can be
written as $ p_0 + \rho_0 = 0 $. This equation of state gives the
vacuum fluid case. Also this solution shows a pressure
distribution which goes to negative infinity as $r$ approaches
zero and density approaches to positive infinity at the centre.
According to Bayin \cite{bay78}, this solution can be used to
represent portions of stars, but can not provide a physically
acceptable model for the entire star.

\begin{figure*}
\begin{center}
\vspace{0.5cm}
\includegraphics[width=0.4\textwidth]{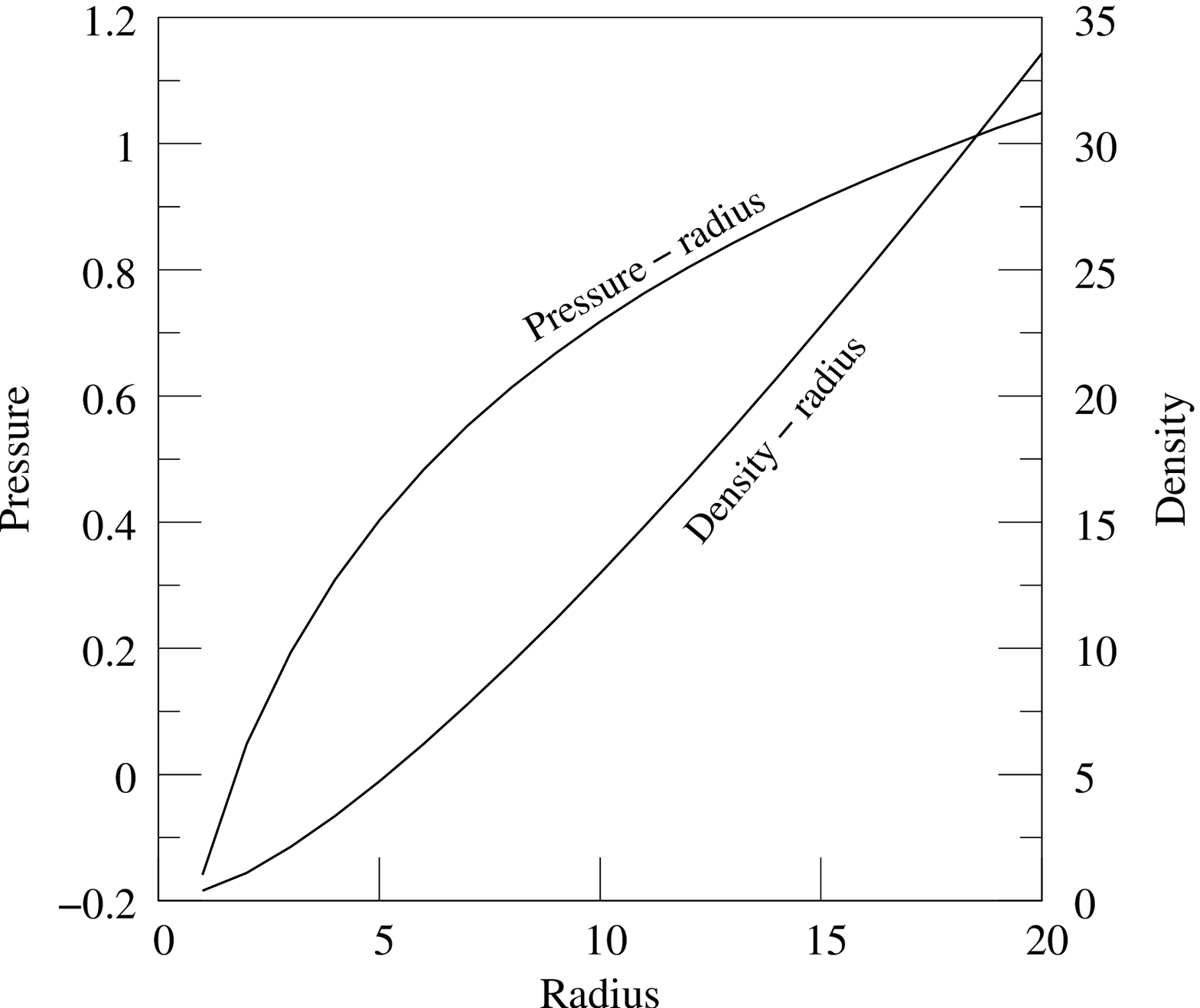}
\includegraphics[width=0.4\textwidth]{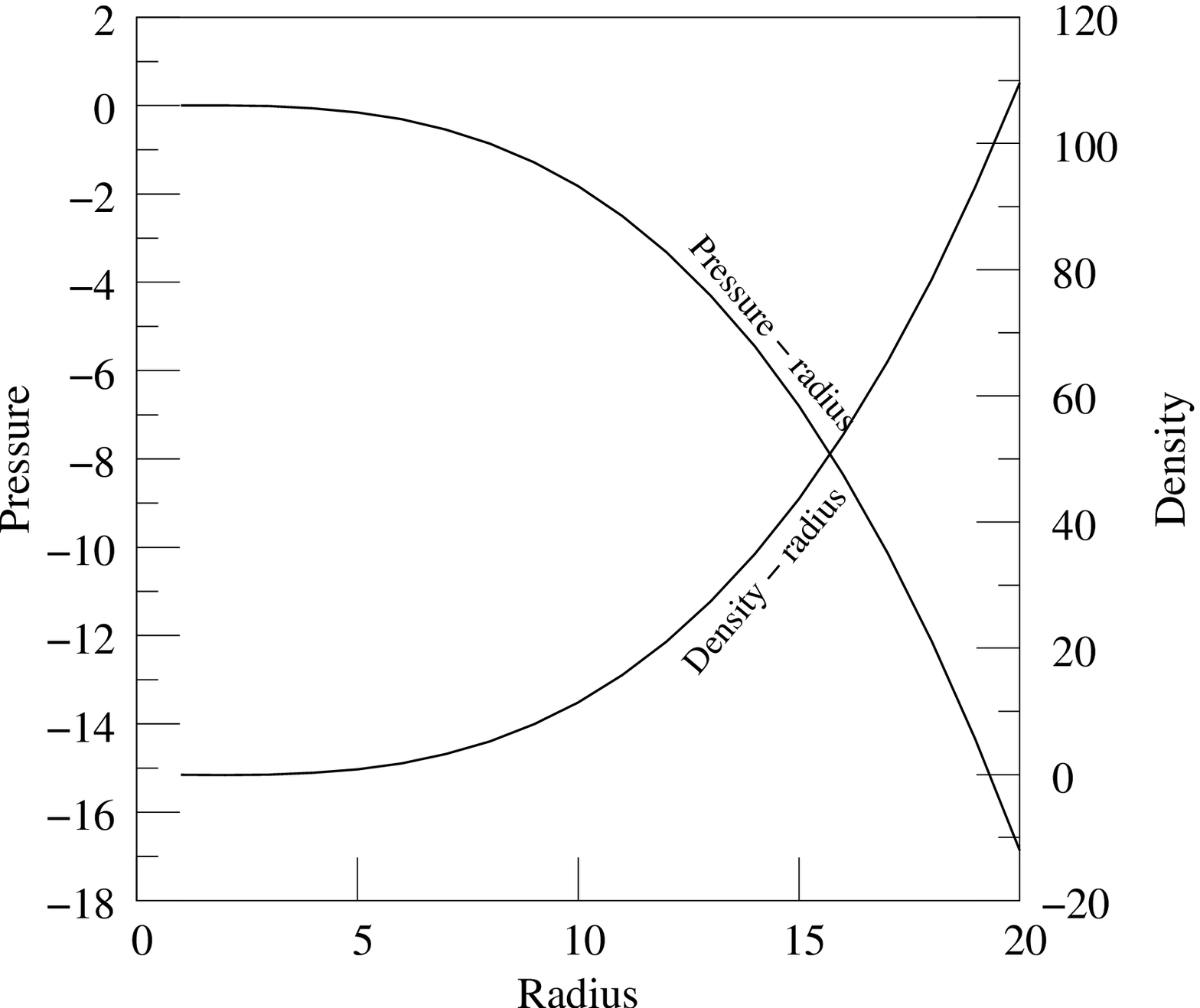}
\caption{Bayin's solutions showing pressure - density - radius plot for
n=1(left) and n=3(right) (Case I). }
\label{fig:bpdr}
\end{center}
\end{figure*}

\noindent{\bf Case I: For $n = 3$}

At the origin, $r=0$, we get $p_c/\rho_c = - 1/3$ which turns into
$3p_c + \rho_c = 0$. This equation of state gives the radiation
case. Also the solution for $p(r)$ shows a pressure distribution
which gives negative infinity pressure at the centre. So it can be
used also to represent a portion of a star. Here mass density is
positive infinity at the origin.

\begin{figure*}
\begin{center}
\vspace{0.5cm}
\includegraphics[width=0.4\textwidth]{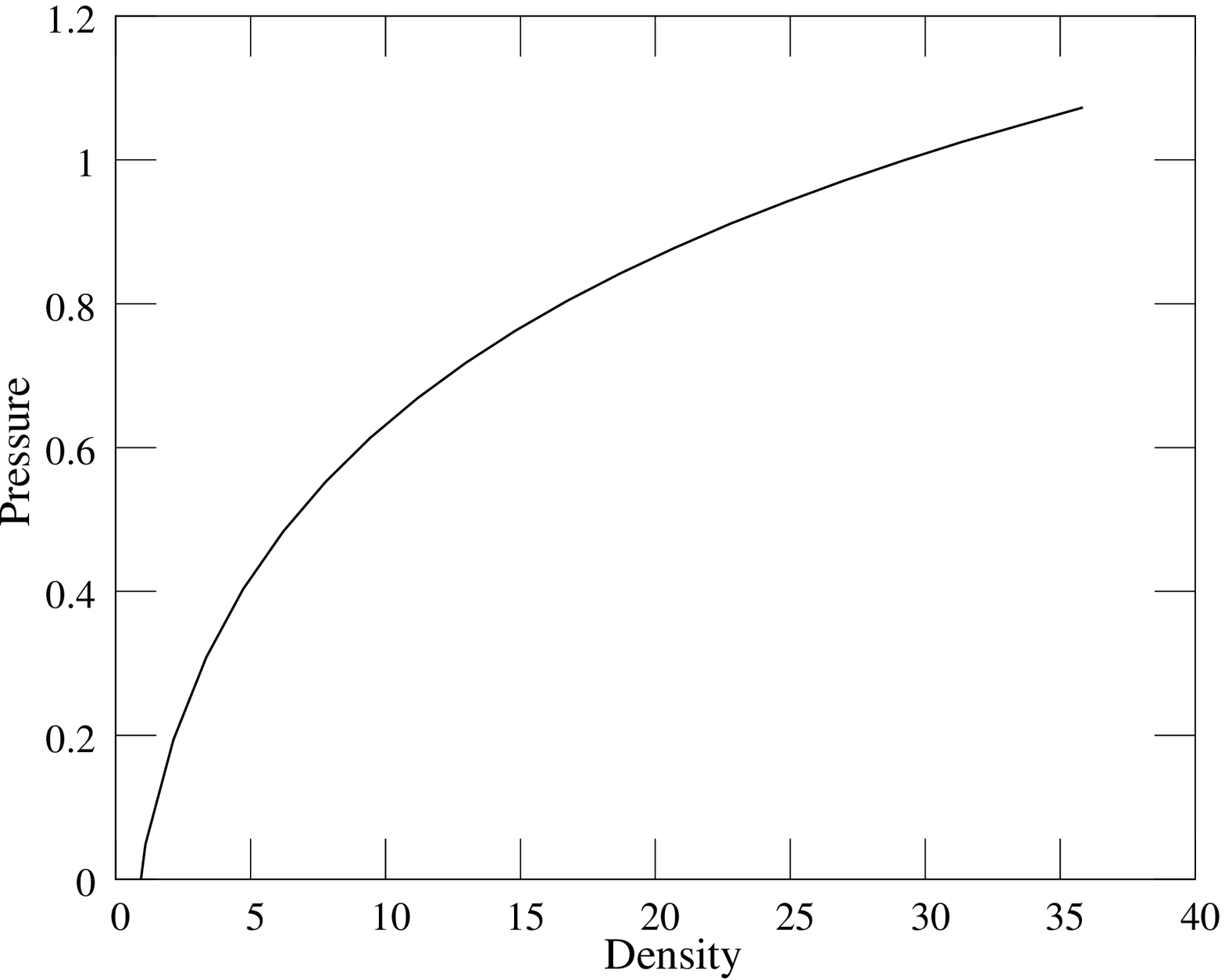}
\includegraphics[width=0.4\textwidth]{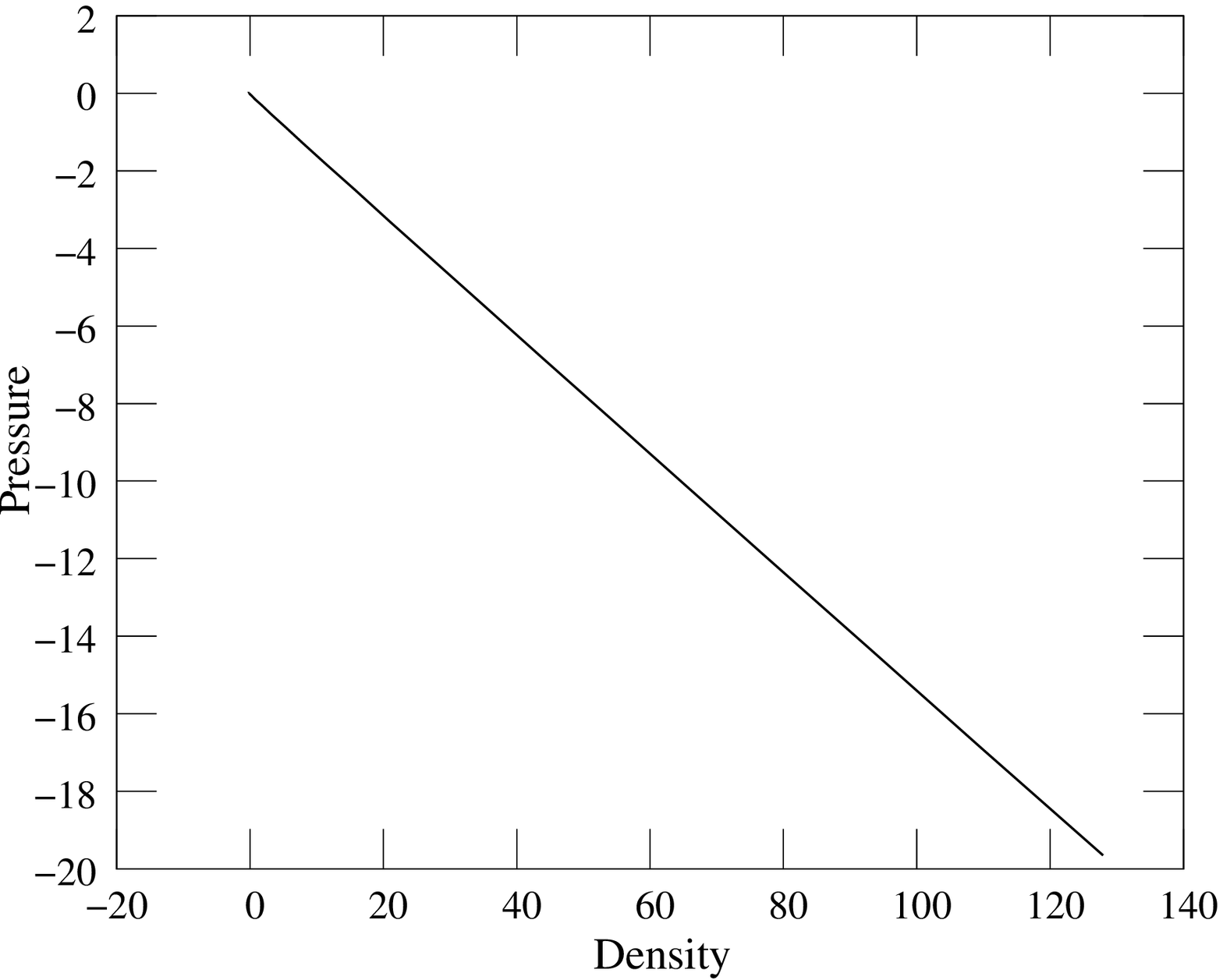}
\caption{The pressure vs density (equation of state) plot for n=1(left) and n=3(right)
for Bayin's solutions (Case I).}
\label{fig:bpd}
\end{center}
\end{figure*}

\noindent{\bf Case II: For $n = 1$}

At the origin it gives $ p_c/\rho_c = -1 $ and we get $p_c +
\rho_c = 0$. This equation of state represents vacuum fluid case.
From the expression of $p(r)$ we see pressure goes to negative
infinity as $r$ approaches zero. Likewise the above case it can be
used to represent a portion of the star where mass density is
positive infinity at the origin.

\noindent{\bf Case II: For $n = 3$}

At the origin, we get the pressure and density as follows,
\begin{eqnarray}
8\pi p_c = \frac{2W_0 + C_1}{W_0 ^2},
\label{bay23pc}
\end{eqnarray}
\begin{eqnarray}
8\pi\rho_c = - \frac{3C_1}{W_0 ^2}.
\label{bay23rhoc}
\end{eqnarray}
Here we see that, at the centre, pressure and density have finite
values. In order to have positive pressure and density at the
origin we must have the following conditions: $ C_1 < 0 $ and $W_0
>|C_1|/2$ where $C_1$ and $W_0$ are constants.

Now, from the Eqs (\ref{bay23pc}) and (\ref{bay23rhoc}) we get
\begin{eqnarray}
\frac{p_c}{\rho_c} = - \frac{2W_0}{3C_1} - \frac{1}{3}.
\label{pcbyrhoc}
\end{eqnarray}
This is exactly the same result as Bayin~\cite{bay78} got in his
non-charge case. Therefore, it is very interesting to note that
for both the cases, corresponding to static neutral-fluid sphere
and static charged-fluid sphere, we get the same result for
central pressure and density.

In Bayin's~\cite{bay78}  paper we see $\rho(r)$ increases as $r$
goes from centre to surface which, indeed, is not to be a
physically reasonable case. But in our expression we have an extra
part $11K^2 r^2/8\pi$ which is arising due to inclusion of charge
where $K$ is a constant. So we can choose suitable values for $K$
such that $\rho(r)$ can be a decreasing function of radius from
centre to surface and hence the solution can provide a physically
valid case. Therefore, for $\rho(r)$ to be a decreasing function
of $r$ from centre to surface we must have, from the Eqs.
(\ref{bay23rho}) and (\ref{bay23rhoc}), the condition
\begin{eqnarray}
\left(\frac{5r^2}{W_0 ^2} - \frac{3C_1}{W_0 ^2} - 11K^2
r^2\right) < \left(- \frac{3C_1}{W_0 ^2}\right),
\end{eqnarray}
which gives $W_0^2 K^2 > 5/11$. Thus, altogether we get three
conditions as follows: $C_1 < 0$, $W_0 > |C_1|/2$ and $W_0^2 K^2
> 5/11$.

Again, for $p(r)$ to be a decreasing function of radius from
centre to surface of the star, we get the condition, from the Eqs.
(\ref{bay23p}) and (\ref{bay23pc}), as
%\begin{widetext}
\begin{eqnarray}
\nonumber
\left[2\left\{C_1 r^2 + W_0^2 - (1 - 2W_0^2 K^2)r^4\right\}^{1/2}
 - r^2 + C_1\right] \\
<  \left[2W_0 + C_1 -3K^2 r^2 W_0^2\right].
\end{eqnarray}
%\end{widetext}
This gives
\begin{eqnarray}
r^2 < \left[\frac{4(C_1 - W_0 + 3K^2 W_0 ^2)}{(5 - 9W_0 ^2 K^2)(1
- W_0 ^2 K^2)}\right].
\end{eqnarray}
The above result is obtained on the condition $5/9 < W_0 ^2 K^2 <
1$. If we apply this condition for radius of the star, for $r =
a$, then we can conclude that through-out the star the pressure
and density have finite values and they are decreasing function
of radius from centre to surface.

Now, we can get the expression for radius of the star considering
$p(r=a)= 0$ at the boundary as
%\begin{widetext}
\begin{eqnarray}
\nonumber
a^2 &=& - \frac{3C_1}{9W_0 ^2 K^2 - 5} \pm
\frac{1}{2}\left[\left(\frac{6C_1}{9W_0 ^2 K^2 - 5}\right)^2 \right. \\
&& \left. - \frac{4(C_1 ^2 -4W_0 ^2)}{(9W_0 ^2 K^2 - 5)(W_0 ^2 K^2 -
1)}\right]^{1/2}.
\end{eqnarray}
%\end{widetext}
By matching interior and exterior solutions on the boundary, $r =
a$, we can evaluate the integration constants $C_1$ and $C_2$ in
terms of mass $m$ and radius $a$ of the star which can be
expressed as follows:
\begin{eqnarray}
C_1 = \left( 1 - 2W_0 ^2 K^2 +\frac{K^2}{W_0 ^2} \right)a^2 -
\frac{2m}{a^3}W_0 ^2,
\end{eqnarray}
%\begin{widetext}
\begin{eqnarray}
\nonumber
C_2 &=& \frac{1}{2}\ln(1 - \frac{2m}{a} + K^2 a^4)
- \frac{1}{(1 - 2W_0 ^2 K^2)^{1/2}} \,\times \\
&&  sin^{-1}\left[\frac{C_1 - 2a^2(1 - 2W_0 ^2
K^2)}{C_1 ^2  + 4W_0 ^2(1 - 2W_0 ^2 K^2)^{1/2}}\right].
\end{eqnarray}
%\end{widetext}
Therefore, after substituting the above value of $C_1$ we can get
a relation between $m$ and $a$ as
%\begin{widetext}
\begin{eqnarray}
\nonumber
m^2 &+& \frac{a^5}{W_0 ^2} \left(2 - W_0 ^2 K^2 - \frac{K^2}{W_0
^2}\right)m  \\
\nonumber
&-& \left(4 - 8W_0 ^2 K^2 + \frac{8K^2}{W_0 ^2} - 10K^4 \right. \\
&& \left. + 7W_0 ^4 K^4+ \frac{K^4}{W_0 ^4}\right)\frac{a^{10}}{4W_0 ^4} +
\frac{a^6}{W_0 ^2} = 0.
\end{eqnarray}
%\end{widetext}
The equation of state for this case which is valid throughout the
sphere is given by
%\begin{widetext}
\begin{eqnarray}
\nonumber
\rho + p &=& \frac{1}{4\pi W_0 ^2} \left[\left\{C_1 r^2 + W_0 ^2 -
( 1 - W_0 ^2 K^2 ) r^4 \right\}^{1/2} \right. \\
&& \left. + 2r^2 - C_1 \right] - \frac{K^2 r^2}{\pi} .
\end{eqnarray}
%\end{widetext}

\noindent{\bf Case III : For $n = 1$}

When $K^2 <5/9$ pressure is negative infinity but density is
positive infinity at the centre of the sphere. Again, if we take
the condition $K^2 > 1$ then the result will be reverse one. We
get, from the boundary condition, the value of the constant $C_3
= (1 + K^2)/m$. Equating pressure to zero at the boundary, $r =
a$, we get the relation between $m$ and $a$ of the star as follows
\begin{eqnarray}
a=\alpha m
\end{eqnarray}
where
\begin{eqnarray}
\alpha = \frac{1}{1 + K^2}\left[1 + \left(1+\frac{1 - K^2}{2K^2 -
1}\right)^{1/2}\right].
\end{eqnarray}
This represents the charge analogue to Bayin's~\cite{bay78} case
III where the relation between the radius and mass was $a = m$ for
the value $K=0$. Again, for our present charged case even putting
$K = 1$ we can recover the same result $a = m$. In this connection
it would be very interesting to note that, from the relation $q(r)
= Kr^n$, for $K = 1$ and $n = 1$ we get the total charge of the
sphere as $q = a$. As a result, $q = m $ which provides the
stability condition regarding a charged fluid sphere. On the
choice of suitable value for $K$ we can get different relation
between $q$ and $m$. A detail study of these type of relation
$a=\alpha m$ have been performed by several worker in different
context~\cite{co78c,fel95yf}.

In Figs. \ref{fig:bpdr} \& \ref{fig:bpd}, we have made a graphical
study of Bayin's solutions for Case I, setting all the
constants to unity, with the only variable being radius, and for
the different $n$ values. This shows a general trend of the nature
of the curve, but fails to match with the realistic stars (compact
and charged). In a future work, we will match conditions to
realistic stars, estimating the exact values of the parameters,
where we can hopefully give a better insight of the solutions. It is also
noteworthy to mention that not all of the Bayin's solutions are
physically feasible to match realistic astrophysical objects, nevertheless
they are of academic importance.

\subsection{Tolman's solutions}

\noindent{\bf Case I: For $n = 0$}

If we now make the specific choice $n=0$ for the parameter $n$
appearing in the above solution set then one get the following
expressions.
\begin{eqnarray}
\rho = \frac{1}{8\pi r^2}\left[\frac{q^2}{a^2}\right],
\end{eqnarray}
\begin{eqnarray}
p = - \frac{1}{8\pi r^2}\left[\frac{q^2}{a^2}\right],
\end{eqnarray}
This gives the equation of state
\begin{eqnarray}
\rho+ p=0
\end{eqnarray}
and the mass-radius relation
\begin{eqnarray}
m = \frac{q^2}{a}.
\end{eqnarray}
We see that it gives EMMM for imperfect fluid case. Now if we look
at the pressure profile, we see it is negative infinity at the
centre and increases from centre to surface. The pressure has a
finite value at the boundary.

\noindent{\bf Case II: For $n = 0.5$ }

We are in favor of this case because of some historical reasons.
This choice was originally done by Tolman~\cite{tol39} himself in
his uncharged version. In this case the gravitational mass becomes
\begin{equation}
m = \frac{3a^2 + 8q^2}{14a}. 
\label{tol38m}
\end{equation}
The equation of state, by virtue of Eqs. (\ref{tol17rho}) and
(\ref{tol18p}), can be written as
\begin{equation}
\rho + p = \frac{1}{4\pi r^2}\left[\frac{n(a^2 - 2q^2)}{(1 + 2n -
n^2)a^2}\right],
\end{equation}
which for the present case reduces to
\begin{equation}
\rho + p = \frac{1}{14\pi r^2}\left[1 - \frac{2q^2}{a^2}\right].
\end{equation}
\begin{eqnarray}
p = \frac{1}{56\pi r^2} \left[1 - \frac{9q^2}{a^2} \right]
\end{eqnarray}
In this case also pressure is negative infinity at the centre. But
if we put the condition $(q/a < \pm 1/3)$ we see pressure becomes
positive throughout the sphere. So, finally we get this case as a
perfect fluid case having positive pressure but not EMMM. At the
boundary, equating pressure to zero, we get $a^2=9q^2$. So, from
Eq. (\ref{tol38m}) one can obtain $m=(5/6)q$ which gives
$q/m>1$.

\noindent{\bf Case III: For $n = 1$ }

For this choice of $n$, the gravitational mass becomes
\begin{equation}
m = \frac{a^2 + 2q^2}{4a}, \label{tol42m}
\end{equation}
whereas equation of state is
\begin{equation}
\rho + p = \frac{1}{8\pi r^2}\left[1 - \frac{2q^2}{a^2}\right].
\end{equation}
\begin{eqnarray}
p = \frac{1}{16\pi r^2} \left[ 1 - \frac{4q^2}{a^2}\right]
\end{eqnarray}
In this case to have positive pressure everywhere in the sphere we
must have the condition $(q/a < \pm 1/2)$. It is also a perfect
fluid case having positive pressure but not EMMM. At the boundary,
equating pressure to zero, we get $a^2=4q^2$. So, from Eq.
(\ref{tol42m}) we have $m=(3/4)q$ which gives $q/m>1$.\\

\noindent{\bf Case IV: For $n =1.5$}

In this case the gravitational mass is
\begin{equation}
m = \frac{3a^2 + 8q^2}{14a}, \label{tol45m}
\end{equation}
and the equation of state becomes
\begin{equation}
\rho + p = \frac{3}{14\pi r^2}\left[1 - \frac{2q^2}{a^2}\right].
\end{equation}
The gravitational mass for this case is the same as that of  $n = 0.5$ case.
\begin{eqnarray}
p = \frac{1}{56\pi r^2} \left[9 - \frac{25q^2}{a^2}\right]
\end{eqnarray}
To have positive pressure throughout the sphere we get the
condition $(q/a < \pm 3/5)$. At the boundary, equating pressure to
zero, one can get $a^2=(25/9)q^2$. So, from Eq. (\ref{tol45m}) we
get $m=(7/10)q$ which gives $q/m>1$.\\

\begin{figure*}
\begin{center}
\vspace{0.5cm}
\includegraphics[width=0.43\textwidth]{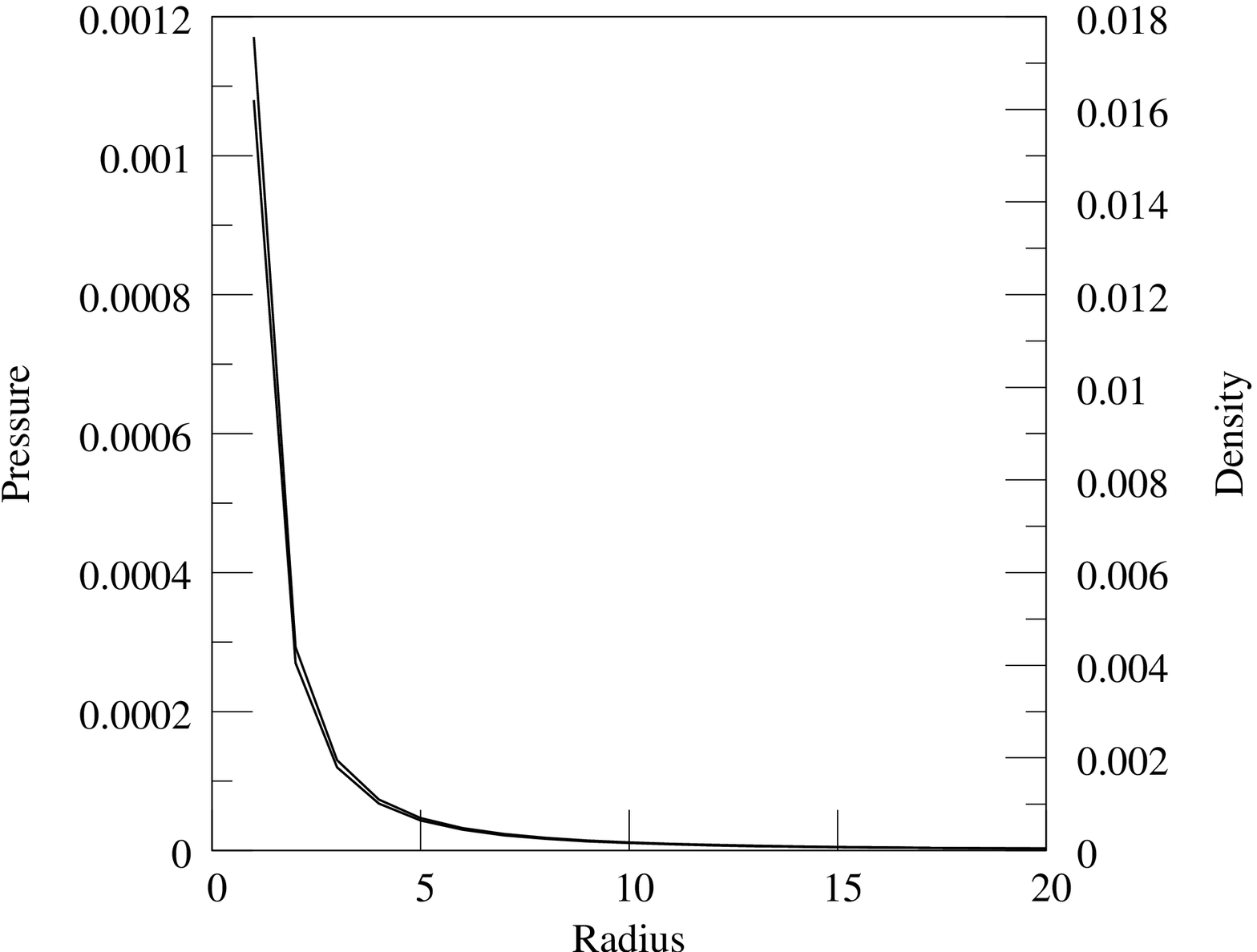}
\includegraphics[width=0.43\textwidth]{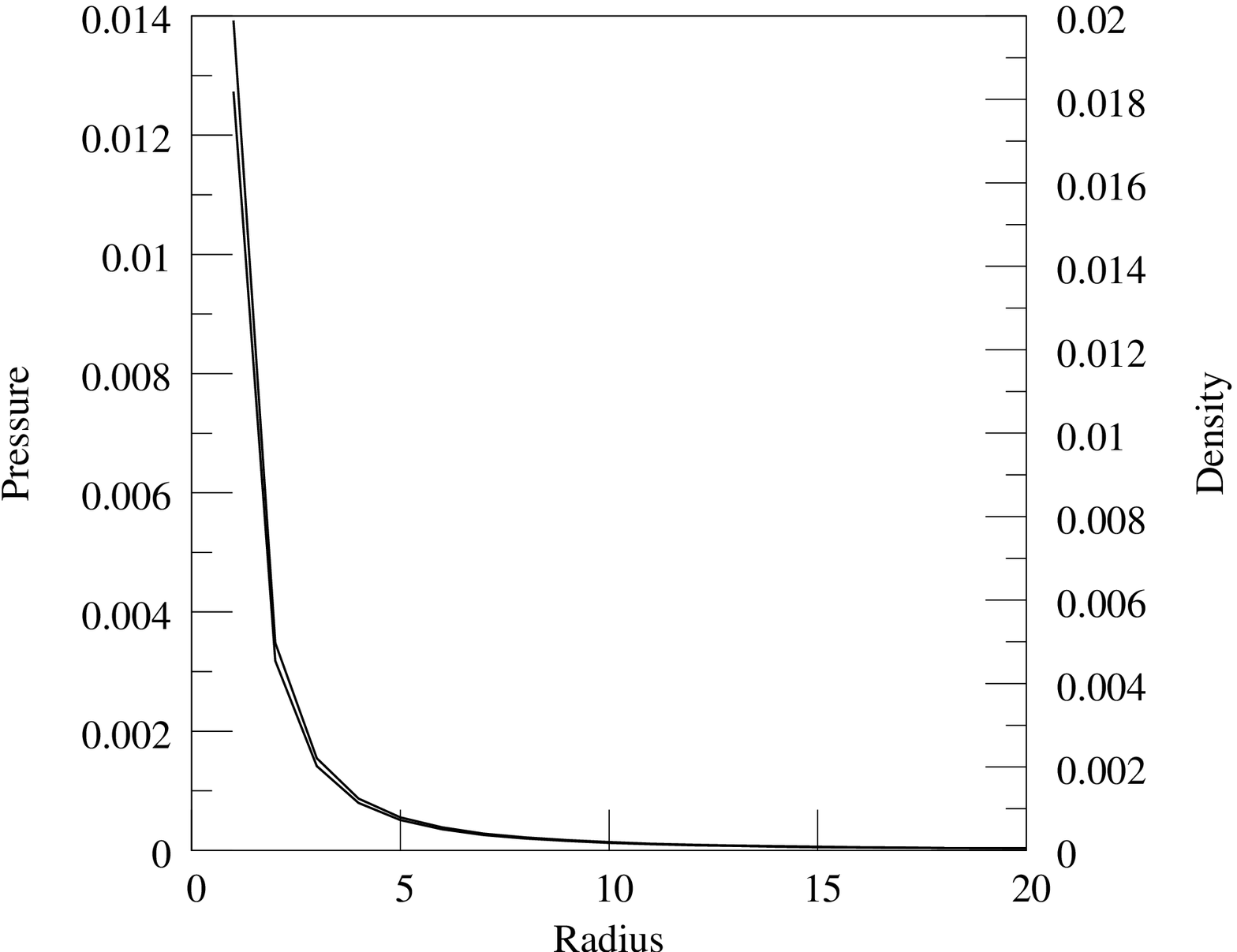}
\includegraphics[width=0.43\textwidth]{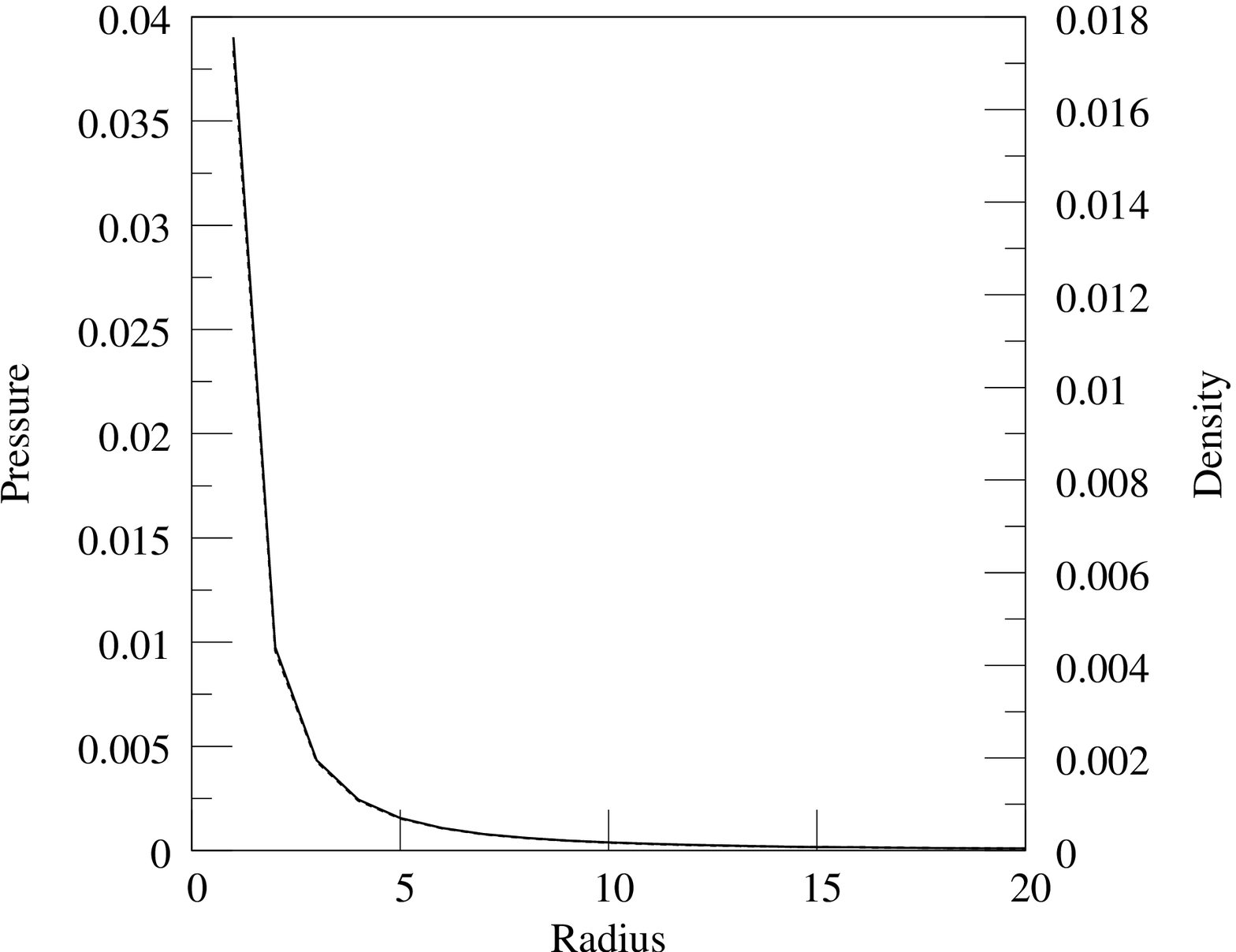}
\includegraphics[width=0.43\textwidth]{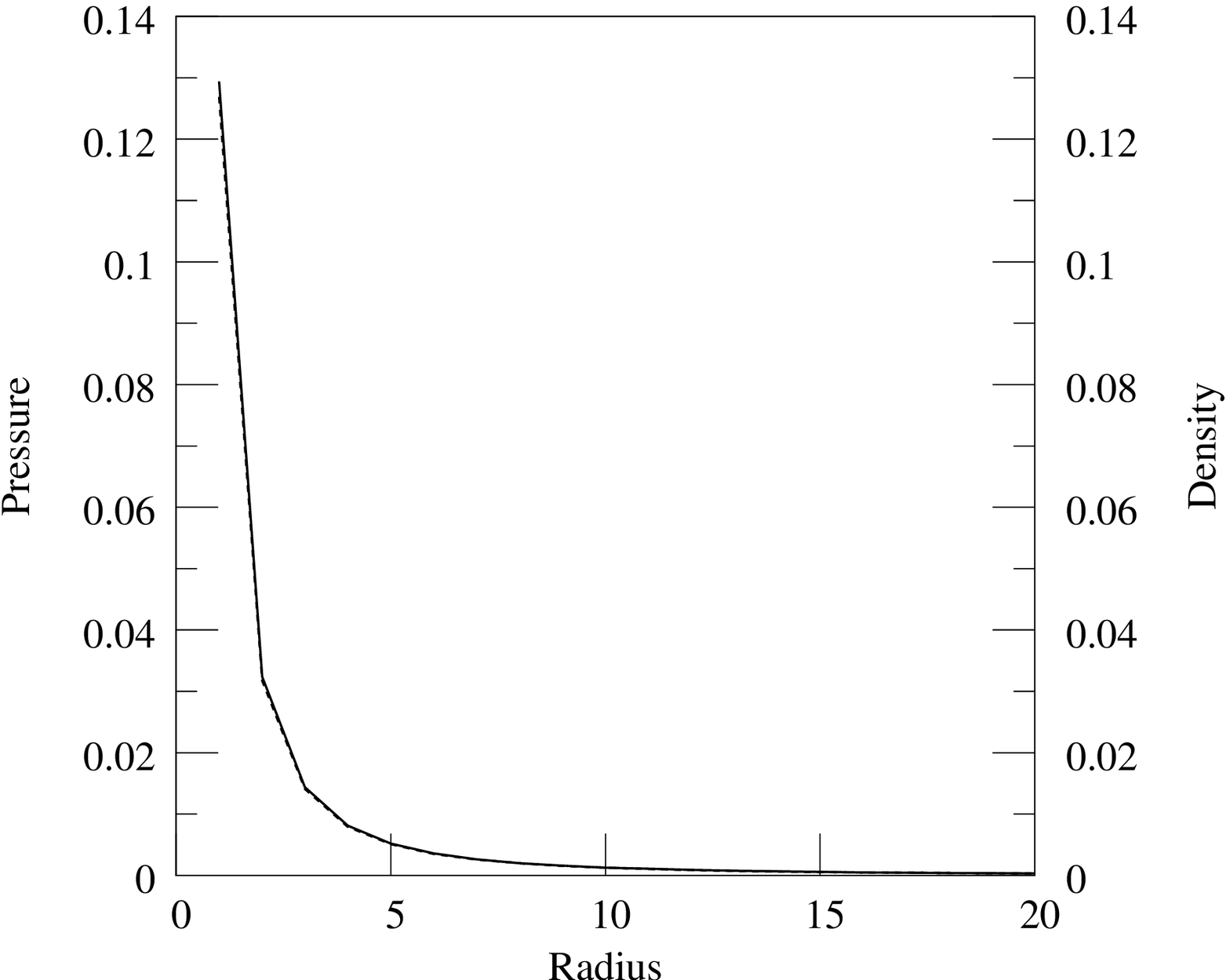}
\caption{Tolman's solutions, showing pressure - density - radius plot for
n=0.5 (top left), n=1.0 (top right), n=1.5 (bottom left) \& n=2.0 (bottom right)
for cases II - V respectively.}
\label{fig:tprd}
\end{center}
\end{figure*}

\begin{figure*}
\begin{center}
\vspace{0.5cm}
\includegraphics[width=0.43\textwidth]{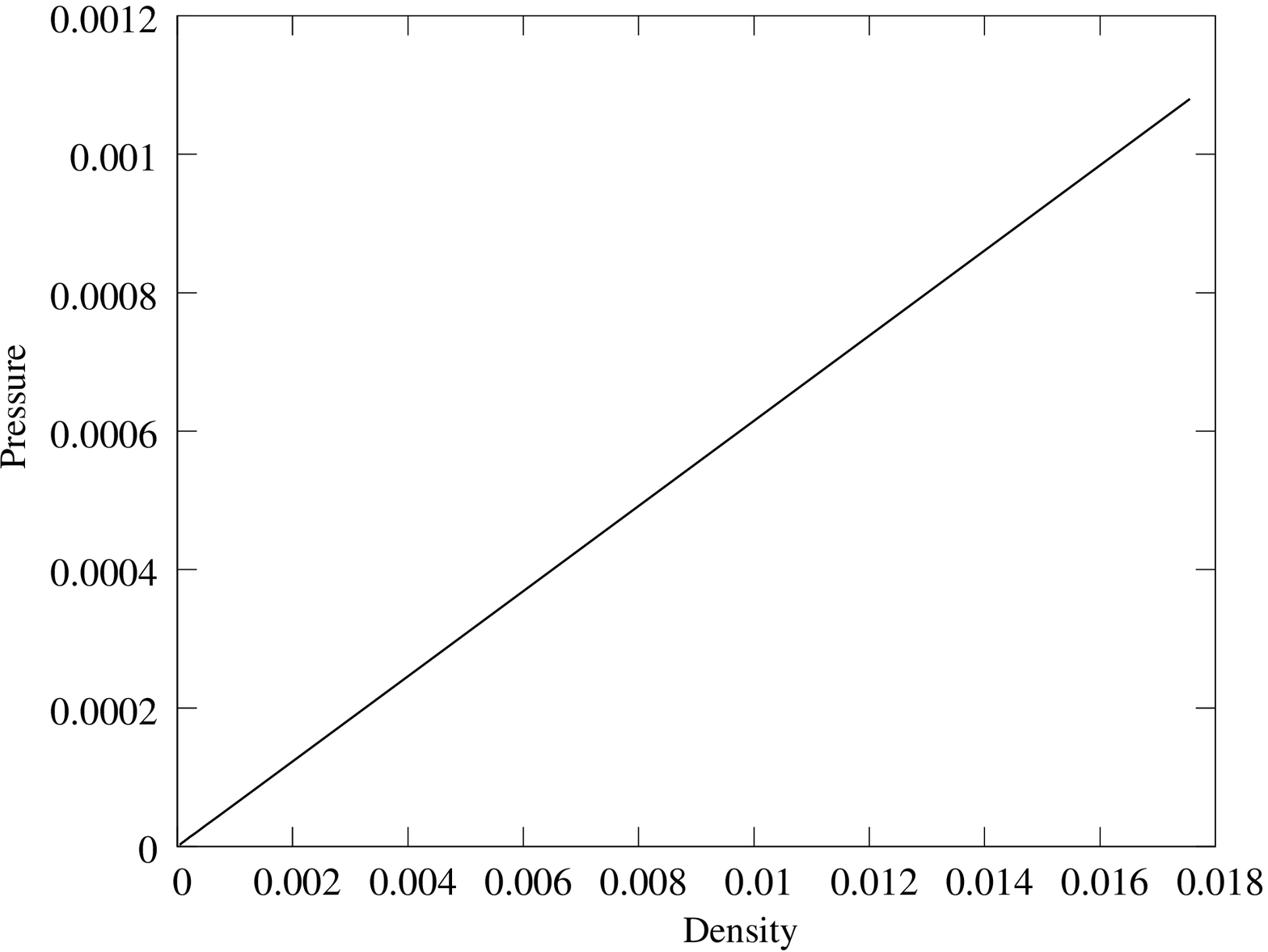}
\includegraphics[width=0.43\textwidth]{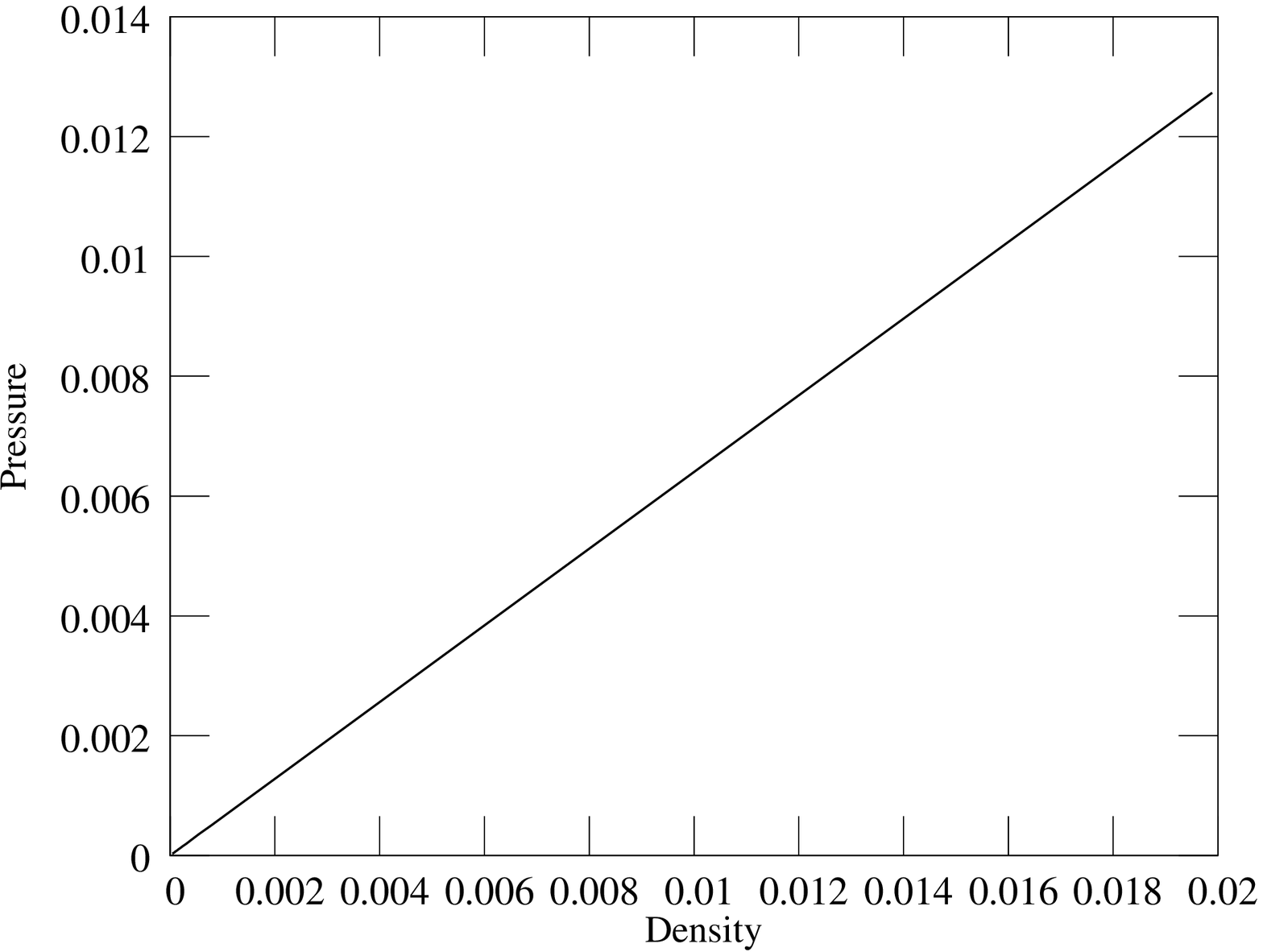}
\includegraphics[width=0.43\textwidth]{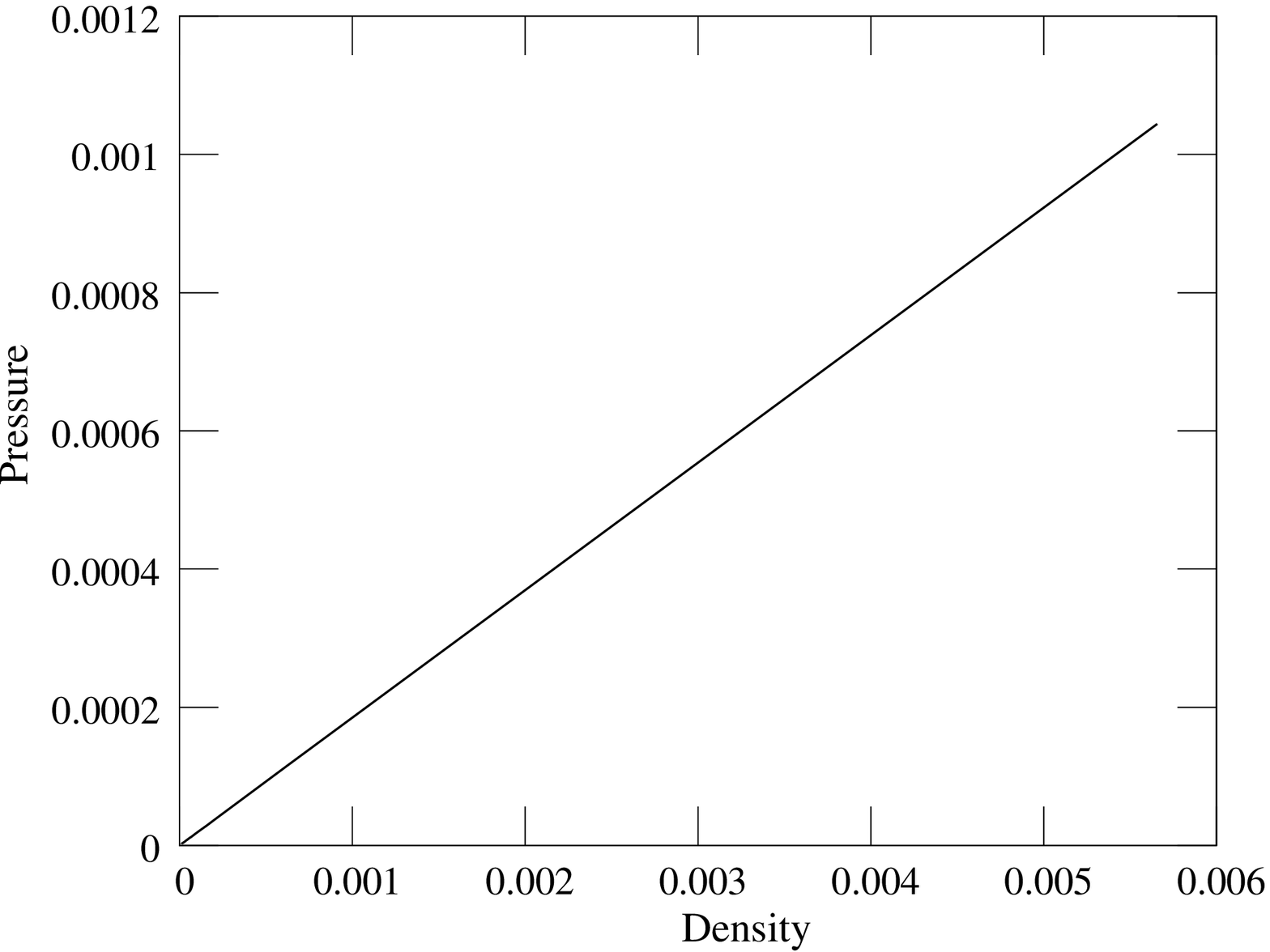}
\includegraphics[width=0.43\textwidth]{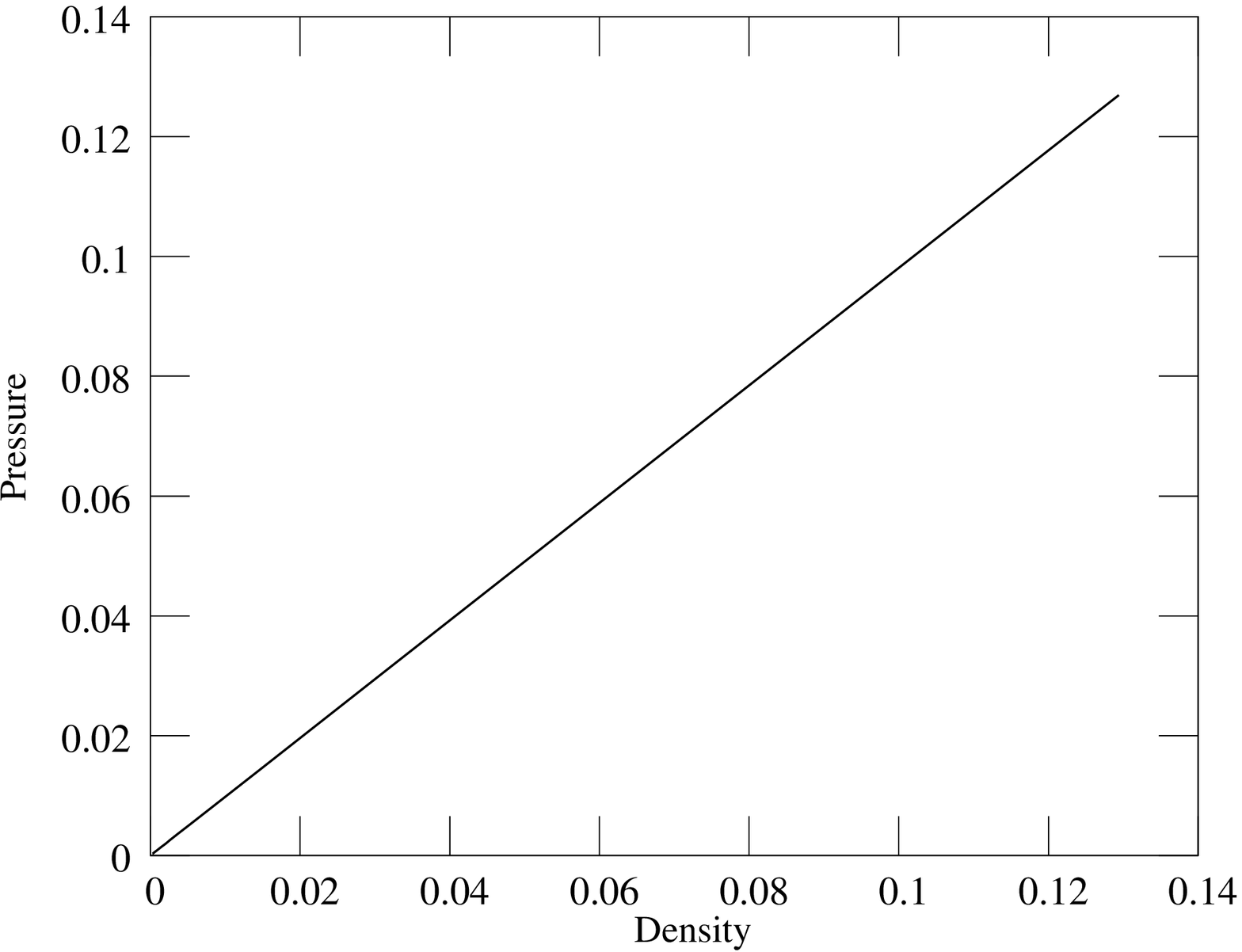}
\caption{Equation of states for n=0.5 (top left), n=1.0 (top right), n=1.5 (bottom left)
\& n=2.0 (bottom right) for Tolman's solutions for cases II - V respectively.}
\label{fig:tpd}
\end{center}
\end{figure*}

\noindent{\bf Case V: For $n = 2$}

The gravitational mass in this case becomes
\begin{equation}
m = \frac{q^2}{a}. \label{tol48m}
\end{equation}
This mass expression is exactly the same as that of the $n = 0$
case of our previous work (Ray \& Das, 2004) which vanishes for the
vanishing electric charge and thus provides `electromagnetic
mass' model (Lorentz, 1904; Feynman et al., 1964). However, the equation of state
for the present situation differs from that of the $n = 0$ case
and is given by
\begin{equation}
\rho + p = \frac{1}{2\pi r^2}\left[1 - \frac{2q^2}{a^2}\right].
\end{equation}
\begin{eqnarray}
p = \frac{1}{8\pi r^2}\left[ 4 - \frac{9q^2}{a^2} \right]
\end{eqnarray}
To have positive pressure throughout the sphere the condition is
$q/a < \pm 2/3$. It is also an interesting case that this
condition gives us it to be a perfect fluid case having positive
pressure and also it gives EMMM. Otherwise we can generalize all
above conditions to one condition which is $q/a < \pm 1/3$ and by
this we get positive pressure for all cases and they are perfect
fluid in nature. At the boundary, equating pressure to zero, we
get $q/a=2/3$. So, from Eq. (\ref{tol48m}) we get $m=(2/3)q$ which
gives $q/m>1$.  In Figs. \ref{fig:tprd} \& \ref{fig:tpd}, we have plotted the
equation of state of the cases described in Tolman's solutions for
the 4 cases (n=0.5, 1.0, 1.5 \& 2) where we have set the uniform condition
$q/a < \pm 1/3$ for all the four cases. For all the above cases
mass density is positive infinity at the centre and decreases from
centre to surface.  It is again to be stressed that the
plots are shown here to give an implication of the nature or behavior
of the solutions, and matching the variables to real units needs some
normalization factors, which is not the primary purpose of this paper
and will be shown elsewhere.

\section{Conclusions}
From discussion in section IV (A), we do not have any physically
viable results for Bayin's solutions for case I, $n = 1,3$ and case II, $n = 1$. All
three cases give negative infinity pressure at the centre. In
this context we would like to mention that idea of negative
pressure is not new in astrophysics, especially in the
realm of cosmology. The equation of state, viz., $\rho + p = 0$
for positive mass density provides a repulsive type negative
pressure \cite{dav84,blo84p,hog84,kai84s}. In that sense, negative
pressure in our solution is not unrealistic. However, for case
II, $n = 3$ we get physically interesting results and hence have
carried out detailed study. At origin, the equation of state is
the same as that Bayin's~\cite{bay78} with neutral case.
Throughout the sphere the equation of state is very complicated.
Moreover, for Bayin's case pressure was a increasing function of
radius from centre to surface and hence according to him it was
not a physically reasonable case. But in our case due to
inclusion of charge we get a condition by which pressure becomes
a decreasing function of radius from centre to surface and thus
becomes a physically acceptable case. Also, this perfect fluid
case provides EMMM with positive pressure. Therefore, it gives an
example which goes in contradiction to Ivanov's~\cite{iva02}
conclusion that ``Electromagnetic mass models are subcases, often
spoiled by negative pressure". For this case, we have also found
out a relation between total mass and radius. For case III,
$n=1$, we get two conditions: one for negative infinity pressure
with positive infinity density and the other one with reverse
features, both at the origin of the sphere. For this case, we get
also a mass-radius relation which gives again the same result in
the absence of charge related to Bayin~\cite{bay78} case. It is
interesting to note that for a particular value of $K=1$, we get
total mass of the spherical system equals the total charge.
Unfortunately, here the fluid pressure becomes negative infinity
in the centre. Otherwise, for other values of $K$ we get finite
values for the ratio of total charge to total mass.

In the Pant \& Sah~\cite{pan79s} cases, which are charge
analogue to Tolman VI type, we have a series of conditions to get
positive as well as finite pressure and density throughout the
sphere. However, from this series of conditions one can easily
arrive at a general condition which is $q/a < \pm 1/3$. One
particular case, for $n = 2$ gives EMMM with positive pressure
under the perfect fluid condition. Again it is an example which
goes in contradiction to Ivanov's conclusion. We also notice that
for $n = 0.5, 1.0, 1.5, 2.0$, total charge to total mass ratio
becomes constant in each case and it is greater than unity. It
reminds the general theorem given by de Felice et al.~\cite{fel99ly}
that {\em if the total electric charge of a
perfect fluid ball is smaller than its total mass, then there is
no regular static configuration having a radius arbitrarily close
to the size of the external horizon}. Therefore, the above four
cases again give support in favor of the stability of the
charged spherical models in connection to normal stars.

\section*{Acknowledgements}
One of the authors (SR1) is thankful to the authority of
Inter-University Centre for Astronomy and Astrophysics, Pune,
India, for providing him Associateship programme under which a
part of this work was carried out. Support under UGC grant (No.
F-PSN-002/04-05/ERO) is also gratefully acknowledged.

%\section*{References}
{}


\begin{thebibliography}{}

\bibitem{buc59} H. A. Buchdhal, Phys. Rev. {\bf 116}, 1027 (1959).

\bibitem{fel95yf} F. de Felice, Y. Yu and J. Fang, Mon. Not. R. Astron. Soc. {\bf 277}, L17 (1995).

\bibitem{sha01mm} R. Sharma, S. Mukherjee and S. D. Maharaj, Gen. Rel. Grav. {\bf 33}, 999 (2001).

\bibitem{iva02} B. V. Ivanov, Phys. Rev. D {\bf 65}, 104001 (2002).

\bibitem{bon65} W. B. Bonnor, Mon. Not. R. Astron. Soc. {\bf 129}, 443 (1965).

\bibitem{ste73} R. Stettner, Ann. Phys. (N.Y.) {\bf 80}, 212 (1973).

\bibitem{gla76} I. Glazer, Ann. Phys. {\bf 101}, 594 (1976).

\bibitem{gla79} I. Glazer, Astrophys. J. {\bf 230}, 899 (1979).

\bibitem{whi81b} P. G. Whitman and R. C. Burch, Phys. Rev. D {\bf 24}, 2049 (1981).

\bibitem{pan79s} D. N. Pant and A. Sah, J. Math. Phys. {\bf 20}, 2537 (1979).

\bibitem{tol39} R. C. Tolman, Phys. Rev. {\bf 55}, 364 (1939).

\bibitem{ray03x} S. Ray, A. L. Esp\'indola, M. Malheiro, J. P. S. Lemos
 \& V. T. Zanchin, Phys. Rev. D {\bf 68}, 084004 (2003); astro-ph/0307262.

\bibitem{ghe05} C. R. Ghezzi, Phys. Rev. D, {\bf 72}, 104017 (2005); gr-qc/0510106.

\bibitem{wey17} H. Weyl, Ann. Physik {\bf 54}, 117 (1917).

\bibitem{maj47} S. D. Majumdar, Phys. Rev. D {\bf 72}, 390 (1947).

\bibitem{pap47} A. Papapetrou, Proc. R. Irish Acad. {\bf 81}, 191 (1947).

\bibitem{pan22} A. Pannekoek, Bull. Astron. Insts. Netherlands, {\bf 1}, 107 (1922).

\bibitem{ros24} S. Rosseland, Mon. Not. R. Astron. Soc. {\bf 84}, 720 (1924).

\bibitem{edd26} A. S. Eddington, The Internal Constitution of the Stars (Cambridge Univ. Press, N. Y., 1926).

\bibitem{cow29} T. G. Cowling, Mon. Not. R. Astron. Soc. {\bf 90}, 140 (1929).

\bibitem{shv71} V. F. Shvartsman, Soviet Physics - JETP {\bf 33}, 475 (1971).

\bibitem{nes01} L. Neslusan, Astron. Astrophys. {\bf 372}, 913 (2001).

\bibitem{bay78} S. S. Bayin, Phys. Rev. {\bf D18}, 2745 (1978).

\bibitem{ray02d} S. Ray and B. Das, Astrophys. Space Sci. {\bf 282}, 635 (2002).

\bibitem{ray04d} S. Ray and B. Das, Mon. Not. R. Astron. Soc. {\bf 349}, 1331 (2004).

\bibitem{ray05d} S. Ray, astro-ph/0409527 (2004).

\bibitem{lor04} H. A. Lorentz, Proc. Acad. Sci. Amsterdam {\bf 6} (1904)
(Reprinted in  Einstein et al., The Principle of Relativity, Dover, INC, p. 24, 1952).

\bibitem{fey64ls} R. P. Feynman, R. B. Leighton and M. Sands, The Feynman Lectures
on Physics (Addison-Wesley, Palo Alto, Vol.~II, Chap. 28, 1964).

\bibitem{co78c} F.I. Cooperstock and V. De La Cruz, Gen. Rel. Grav. {\bf 9}, 835 (1978).

\bibitem{dav84} C. W. Davies, Phys. Rev. {\bf D30}, 737 (1984).

\bibitem{blo84p} J. J. Blome and W. Priester, Naturwissenshaften {\bf 71}, 528 (1984).

\bibitem{hog84} C. Hogan, Nature {\bf 310}, 365 (1984).

\bibitem{kai84s} N. Kaiser and A. Stebbins, Nature {\bf 310}, 391 (1984).

\bibitem{fel99ly} F. de Felice, S. Liu and Y. Yu, Class. Quan. Grav. {\bf 16}, 2669 (1999).

\end{thebibliography}
\end{document}